\documentclass[sigplan,screen,nonacm]{acmart}

\usepackage{xcolor}
\definecolor{codegreen}{rgb}{0,0.6,0}

\usepackage{cancel}

\usepackage{listings}
\renewcommand\O[1]{$\mathcal{O}(#1)$}

\setcopyright{cc}
\setcctype{by}
\acmDOI{10.1145/3677999.3678282}
\acmYear{2024}
\copyrightyear{2024}
\acmISBN{979-8-4007-1102-2/24/09}
\acmConference[Haskell '24]{Proceedings of the 17th ACM SIGPLAN International Haskell Symposium}{September 6--7, 2024}{Milan, Italy}
\acmBooktitle{Proceedings of the 17th ACM SIGPLAN International Haskell Symposium (Haskell '24), September 6--7, 2024, Milan, Italy}
\acmSubmissionID{icfpws24haskellmain-p94-p}
\received{2024-06-03}
\received[accepted]{2024-07-05}

\begin{document}

\title{Liquid Amortization: Proving Amortized Complexity with LiquidHaskell (Functional Pearl)}

\author{Jan van Brügge}
\orcid{0000-0003-1560-7326}
\affiliation{%
  \institution{Heriot-Watt University}
  \city{Edinburgh}
  \country{United Kingdom}
}
\email{jsv2000@hw.ac.uk}


\begin{abstract}
Formal reasoning about the time complexity of algorithms and data structures is usually done in interactive theorem provers like Isabelle/HOL~\cite{isabelle}. This includes reasoning about \textit{amortized} time complexity which looks at the worst case performance over a \textit{series} of operations. However, most programs are not written within a theorem prover and thus use the data structures of the production language. To verify the correctness it is necessary to translate the data structures from the production language into the language of the prover. Such a translation step could introduce errors, for example due to a mismatch in features between the two languages.

We show how to prove amortized complexity of data structures directly in Haskell using LiquidHaskell~\cite{liquidhaskell}. Besides skipping the translation step, our approach can also provide a didactic advantage. Learners do not have to learn an additional language for proofs and can focus on the new concepts only. For this paper, we do not assume prior knowledge of amortized complexity as we explain the concepts and apply them in our first case study, a simple stack with multipop. Moving to more complicated (and useful) data structures, we show that the same technique works for binomial heaps which can be used to implement a priority queue. We also prove amortized complexity bounds for Claessen's version of the finger tree~\cite{fingertrees_new}, a sequence-like data structure with constant-time cons/uncons on either end. Finally we discuss the current limitations of LiquidHaskell that made certain versions of the data structures not feasible.
\end{abstract}



\maketitle

\section{Introduction}

Formally proving properties of production code is becoming more and more common. For safety critical code that is not allowed to "go wrong" as well as code that is fundamental and reused very often, like core libraries, bugs can have a widespread impact. Proving that the code behaves according to the specification can provide an extra layer of confidence. Fundamental data structures are one of the most important parts of a programming language ecosystem and a lot of production code depends on them. In languages like Haskell this comes mostly in the form of functional, persistent data structures like the ones described in Okasaki's book on purely functional data structures~\cite{okasaki}.

Usually formal proofs are done in an external theorem prover (compare e.g.~\cite{complexity_coq} or~\cite{complexity_isabelle}), requiring an explicit translation step from the production language to the language the theorem prover uses. Such a translation step could introduce subtle differences or bugs if this translation is done manually. Additionally, the target language may not support certain features present in the source language (e.g. polymorphic recursion in data types, see section~\ref{sec:fingertree}). In those cases, the missing features have to be emulated in the target language, again opening the door for small and subtle errors. Because the code that is actually used in the formal proofs differs from the source code that is running in production, it is often hard to reconnect the proofs back to the original production code.

Another possibility is using LiquidHaskell~\cite{liquidhaskell}. It is a refinement type~\cite{refinement_types} checker plugin\footnote{Before 2020, LiquidHaskell was implemented as a standalone executable; it now being a GHC plugin~\cite{lh_plugin} drastically improves usability} for GHC, the main Haskell compiler used in industry. It allows specifying additional constraints on types, e.g. "integers greater than zero" instead of just "integers". These constraints are propagated through the code and checked by an external SMT solver. With a handful of proof combinators defined by LiquidHaskell, it is possible to use Haskell directly as a theorem prover, enabling intuitive equational reasoning within the language~\cite{tpfa}. This makes it easy to keep the implementation and its respective proof in sync, as any failure to do so will result in a compiler error. LiquidHaskell errors are also integrated with the normal Haskell IDE tooling, so the user can see proof failures directly in their editing environment. Another benefit is that the reader or author of a proof does not need to learn an additional language, the production language is the same as the proof language.

In this paper, we:
\begin{itemize}
\item{Assume no prior knowledge of amortized complexity as we recap the theoretical background (section~\ref{sec:complexity}).}
\item{Show the reasoning infrastructure that is needed to prove amortized complexity using LiquidHaskell (section~\ref{sec:liquidhaskell}).}
\item{Apply this technique to several case studies: a simple \textit{stack} (section~\ref{sec:stack}), \textit{binomial heaps} (section~\ref{sec:binomal_heap}) and \textit{finger trees} (section~\ref{sec:fingertree}).}
\item{Discuss the limitations of LiquidHaskell for such proofs compared with proper theorem provers (section~\ref{sec:conclusion}).}
\end{itemize}

\section{Amortized complexity}\label{sec:complexity}

It is common practise to investigate how many units of time an operation on a data structure takes in relation to the number of elements in the data structure. As the performance can vary wildly for the same operation depending on the state of the data structure, usually the worst case for each operation is used as a metric and specified using \textit{Landau notation} ("big O"). For example, a linear search\footnote{walking through a list element by element, checking if the current element is the one that is searched for} takes $1$ unit of time if the searched element is at the front, while it takes $n$ units of time if it is the last element (where $n$ is the number of elements in the list). Thus we would say that linear search is in the complexity class of \O{n} because it will not take longer than $c \cdot n$ units of time for some fixed constant $c$.

But often this worst case is too pessimistic. When an operation changes the data structure, for example by removing elements, consecutive operations might be cheaper. Or to use an expensive operation it might be necessary to have a certain amount of cheap operations. In such cases one uses \textit{amortized complexity}, which does not only look at a single operation in isolation, but at chains of operations.

As a running example we will use a stack with \textit{push} and \textit{multipop} operations. The former allocates a new element and a reference to the rest of the stack and thus takes a constant amount of time. The latter removes the first $k$ elements from the stack. In the worst case, it removes all the elements, making \texttt{multipop} \O{n}. One can already see that this is too pessimistic, because after removing all elements from the stack, the next \texttt{multipop} will do nothing. In fact, to be able to remove an element, it has had to be pushed to the stack at some point, so we can spread out the cost of \texttt{multipop} over those pushes.

There are multiple techniques to prove amortized complexity; we present the two most popular ones. Note that both of these techniques only apply in a non-shared setting where intermediate results are not reused. If we were allowed to keep a reference to the original data structure after an operation, we could do the same expensive operation several times without "paying" for it. Some data structures can exploit laziness to have the same (good) amortized complexity also in a shared setting, however formally verifying this would require explicit reasoning about the lazy semantics of the language and thus massively complicate the analysis. Note that in a non-shared setting the amortized complexity bounds apply regardless of whether the language is lazy or strict. Laziness may cause the data structure to do less work -- and thus have a better complexity -- but as an upper bound the amortized complexity is still valid.

\subsection{Banker's method}\label{sec:banker}

The banker's method uses a "bank" account that can store time units. Operations have to pay their actual runtime cost and may additionally deposit into or withdraw from the account. The sum of the actual cost and the deposit/withdrawal is the \textit{amortized cost} of the operation. The idea is that cheap operations pay extra time units, making their amortized cost higher than their actual cost. The goal is to keep the sum of the actual cost and the extra deposit in the same (good) complexity class as just the actual cost. Expensive operations on the other hand can use the stored time to pay for their actual cost, hopefully putting them in a better complexity class. In our example a \texttt{push} has to pay one time unit as actual cost and it will also pay one extra time unit into the bank account. The sum of the amortized cost and actual cost still constant, so \texttt{push} stays in \O{1}. Because for every element pushed we deposit one extra time unit into the bank, the account always contains as many time units as there are elements on the stack. As \texttt{multipop} cannot remove more elements than there are on the stack, withdrawing $k$ time units to pay for the actual cost will never make the account become negative. Because withdrawing from the bank account is enough to pay for the complete actual cost of \texttt{multipop} regardless of the choice of $k$, its \textit{amortized complexity} is in \O{1}.

\subsection{Physicist's method}\label{sec:physicist}

While the banker's method is easy to visualize, it is hard to use in a formal proof because the bank account acts as state between the operations. A simpler method is the physicist's method, which does not need such an account. The idea is to define a potential $\Phi$ for the data structure. This potential is defined to be $0$ for an empty data structure and $\ge 0$ for all other states. Intuitively, the potential represents how much time needs to be saved up to pay for expensive operations. We then define the amortized cost of an operation as $c + \Phi(h') - \Phi(h)$ where $c$ is the actual time cost of the operation, $h$ is the data structure before the operation and $h'$ is the data structure afterwards. Looking at a chain of operations $1, 2, ..., n$, the sum of the amortized time is an overestimation of the actual cost by $\Phi(h_n)$ and thus an upper bound on the actual cost (see equation~\ref{eq:sum}, remember that by definition $\Phi(h_0) = 0$ and $\Phi(h_n) \ge 0$).

\begin{gather}
\nonumber (c_1 + \cancel{\Phi(h_1)} - \Phi(h_0)) + (c_2 + \cancel{\Phi(h_2)} - \cancel{\Phi(h_1)}) \\
\nonumber    + ... + (c_n + \Phi(h_n) - \cancel{\Phi(h_{n - 1})}) = \\
  c_1 + c_2 + ... + c_n + \Phi(h_n) - \Phi(h_0) \label{eq:sum}
\end{gather}

For our stack we can define the potential to be the height of the stack. This fulfils the requirements as the empty stack has a potential of zero and all other states have a potential greater or equal to zero. For the actual proof of this example see the first case study in section~\ref{sec:stack}.

\section{LiquidHaskell for theorem proving}\label{sec:liquidhaskell}

Usually LiquidHaskell is used to add pre- and postconditions to functions with the help of refinement types~\cite{refinement_types}. The refinements are additional constraints on top of normal Haskell data types. These constraints are then checked by an external SMT solver. By default LiquidHaskell uses the z3~\cite{z3} solver, but it also supports cvc4~\cite{cvc4}. For example, it allows us to specify that the caller of a division function has to ensure that the denominator is not zero.

\begin{lstlisting}
{-@ div' :: Int -> { x:Int | x /= 0 } -> Int @-}
div' :: Int -> Int -> Int
div' = div
\end{lstlisting}

LiquidHaskell annotations are multiline Haskell comments delineated by @ signs, so that the code still compiles even without LiquidHaskell. It also allows the user to name argument types, so they can be referred to later. Alternatively there exists a Quasiquoter that would allow us to get rid of the duplication between the Haskell and the LiquidHaskell type signatures.

In LiquidHaskell, all functions have to terminate. Usually a suitable termination metric is automatically deduced, but in some more complex cases, we can also manually supply this metric. From this termination metric, LiquidHaskell is able to generate an induction principle for the function that can be used in a proof. For example to prove that the length of a list is never negative, LiquidHaskell uses the induction hypothesis to assume that the length of some list \texttt{xs} is not negative and uses that to prove \texttt{length (x:xs)} is not negative:

\begin{lstlisting}
{-@ measure length @-}
{-@ length :: [a] -> { x:Int | x >= 0 } @-}
length :: [a] -> Int
length [] = 0 -- trivial, 0 >= 0
length (x:xs) = 1 + length xs
    -- by induction: length xs >= 0
    -- thus 1 + length xs >= 0
\end{lstlisting}

Refinements can use equality (written as \texttt{==}) and comparison without the need for the \texttt{Eq} or \texttt{Ord} typeclasses. The reason is that in refinments these operators are translated to the builtin operators of the SMT theory and thus cannot be customized. To use functions in the refinements, one has to \textit{reflect}~\cite{reflection} them to the logic level. For a limited class of functions called \textit{measures}, this can be done while retaining full automatic reasoning. A measure is a function that takes only one argument which is an algebraic data type, has one equation per constructor of this data type and which only uses arithmetic functions and other measures in its right hand sides. Other functions still can be used in refinements when annotated with \texttt{reflect}, but the SMT solver is not able to check those refinements automatically. The user always has to provide the proofs themselves using LiquidHaskell's ability to do complex equational machine-checked reasoning~\cite{tpfa}.

To facilitate these proofs, LiquidHaskell provides a set of proof combinators and a \texttt{Proof} type. This type is just a type alias for the normal Haskell unit type that is refined by the property that needs to be proved. The reasoning comes from the \texttt{===} operator that asserts that both sides of the equality have the same refinement. We can bring other facts in scope using the \texttt{?} operator. The angle brackets used in its refinement type are used to explicity quantify over two refinements. This will make the facts "visible" in the type signature and thus make them available to the SMT solver. A proof can be finished by the \texttt{***} operator that casts a chain of equalities to the proof type (type slightly simplified here).

\begin{lstlisting}
type Proof = ()

trivial :: Proof
trivial = ()

infixl 3 ===
{-@ (===) :: x:a -> y:{a | y == x}
  -> {v:a | v == x && v == y} @-}
(===) :: a -> a -> a
_ === y  = y

infixl 3 ?
{-@ (?) :: forall a b
  <pa :: a -> Bool, pb :: b -> Bool>.
  a<pa> -> b<pb> -> a<pa> @-}
(?) :: a -> b -> a
x ? _ = x

data QED = QED
infixl 3 ***
{-@ assume (***) :: a -> p:QED -> { true } @-}
(***) :: a -> QED -> Proof
_ *** _ = ()
\end{lstlisting}

To illustrate we prove that the length of \texttt{append} applied to two lists is the sum of the lengths of the two lists. Because \texttt{append} takes more than one argument, it is not a measure and thus we need to do the proof manually.

For this proof, we need to do a case distinction on the first argument. This is done by pattern matching on it like in a normal Haskell function. The rest of the proof is done by applying the definitions of the used functions, arithmetic equalities and induction:

\begin{lstlisting}
{-@ reflect append @-}
append :: [a] -> [a] -> [a]
append [] ys = ys
append (x:xs) ys = x:(append xs ys)

{-@ lengthP :: xs:[a] -> ys:[a] ->
  { length (append xs ys) ==
      length xs + length ys }
@-}
lengthP :: [a] -> [a] -> Proof

-- base case
lengthP [] ys = length (append [] ys)
  -- Use definition of append [] _
    === length ys
    === 0 + length ys
  -- use definition of length []
    === length [] + length ys
    *** QED

-- recursive case
lengthP (x:xs) ys =
    length (append (x:xs) ys)
  -- Use definition of append
    === length (x:(append xs ys))
  -- Use definition of length
    === 1 + length (append xs ys)
  -- Use induction with the ? operator
    ? lengthP xs ys
    === 1 + length xs + length ys
  -- Use definition of length
    === length (x:xs) + length ys
    *** QED
\end{lstlisting}

This proof is quite lengthy, because we have to manually unfold definitions. LiquidHaskell also has a feature called \textit{proof by logical evaluation (ple)}. With this feature turned on, definitions are automatically unfolded several times for the SMT solver. Of course doing that for every definition might result in a performance hit, so it is also possible to selectively enable this feature on a per definition basis by tagging it with the \texttt{automatic-instances} flag. With \textit{ple} enabled, the proof boils down to doing the case distinction and specifying the induction step:

\begin{lstlisting}
{-@ automatic-instances lengthP @-}
{-@ lengthP :: xs:[a] -> ys:[a] -> {
  length (append xs ys) == length xs + length ys
} @-}
lengthP :: [a] -> [a] -> Proof
lengthP [] _ = trivial
lengthP (_ : xs) ys = trivial ? lengthP xs ys

\end{lstlisting}

\section{Case study: Stack with Multipop}\label{sec:stack}

To continue with the example from section~\ref{sec:complexity}, we prove that $multipop$ has an amortized complexity of \O{1}. This data structure and its operations is defined below. We already add a precondition to assert that $k$ is not negative with the help of a LiquidHaskell type synonym.

\begin{lstlisting}
{-@ type Nat = { x:Int | x >= 0 } @-}

data Stack a = Empty | Elem a (Stack a)

{-@ reflect push @-}
push :: a -> Stack a -> Stack a
push x s = Elem x s

{-@ reflect multipop @-}
{-@
  multipop :: Nat -> Stack a -> ([a], Stack a)
@-}
multipop :: Int -> Stack a -> ([a], Stack a)
multipop _ Empty = ([], Empty)
multipop 0 s = ([], s)
multipop n (Elem x s) =
  let (xs, s') = multipop (n - 1) s
  in (x:xs, s')
\end{lstlisting}

To formally prove the amortized complexity with the physicist's method, we first need to define a potential $\Phi$ and a timing function for each of our operations that gives us the actual time cost of that operation. As seen in section~\ref{sec:physicist}, we use the height of the stack as our potential.

The timing functions tell us the cost of every operation. Their definition follow the recursive structure of the operations. We could use TemplateHaskell to automatically derive these timing functions from the original operations but for clarity we will define them manually throughout this work.

\begin{lstlisting}
{-@ reflect phi @-}
{-@ phi :: Stack a -> Nat @-}
phi :: Stack a -> Int
phi Empty = 0
phi (Elem _ s) = 1 + phi s

{-@ reflect pushT @-}
{-@ pushT :: a -> Stack a
      -> { x:Int | x >= 1 } @-}
pushT :: a -> Stack a -> Int
pushT _ _ = 1 -- no recursion in
               -- original definition

{-@ reflect multipopT @-}
{-@ multipopT :: Nat -> Stack a
     -> { x: Int | x >= 1 } @-}
multipopT :: Int -> Stack a -> Int
multipopT _ Empty = 1
multipopT 0 _ = 1
multipopT n (Elem _ s) = 1 + multipopT (n - 1) s
\end{lstlisting}

First, we will prove the complexity of \texttt{push}. While the amortized cost is in the same complexity class as the actual cost, we need to make sure that our potential function does not cause those two to diverge. LiquidHaskell does not allow quantifiers in refinements, because for functions other than measures, instantiating quantifiers in the SMT solver has unpredictable performance~\cite{lh_quantifiers}. This means we have to guess an upper limit for our amortized runtime. In section~\ref{sec:banker}, we paid two time units for a push to amortize the pops, so we will use $2$ as our guess:

\begin{lstlisting}
import Language.Haskell.Liquid.ProofCombinators

{-@ pushP :: x:a -> s:Stack a ->
  { pushT x s + phi (push x s) - phi s <= 2 }
@-}
pushP :: a -> Stack a -> Proof
pushP x s = pushT x s + phi (push x s) - phi s
-- Use definition of push and pushT
  === 1 + phi (Elem x s) - phi s
-- Use definition of phi (Elem _ s)
  === 1 + (1 + phi s) - phi s
-- phi s gets canceled, 1 + 1 <= 2
  *** QED
\end{lstlisting}

To prove the complexity of \texttt{multipop}, we will need to distinguish between the two base cases and the recursive case. With LiquidHaskell this is just a normal pattern match. Leaving the recursive case undefined for now, proving the base cases is very similar to \texttt{push} so we use \textit{ple} to automate the proofs:

\begin{lstlisting}
{-@ automatic-instances multipopP @-}
{-@ multipopP :: k:Nat -> s:Stack a ->
  { multipopT k s + phi (snd (multipop k s))
      - phi s <= 2 }
@-}
multipopP :: Int -> Stack a -> Proof
multipopP 0 _ = trivial
multipopP _ Empty = trivial
\end{lstlisting}

To avoid cluttering the proof of the recursive case, we use an \textit{as pattern} to alias \texttt{Elem x s} to \texttt{xs}. The first steps are again applying the definitions of the used functions and simplifying the result. In the end we use the \texttt{?} operator to add the induction hypothesis to the proof step and thus complete the proof. We again use \textit{ple} here to unfold the definition of \texttt{snd} and \texttt{multipop} at once.

\begin{lstlisting}
multipopP k xs@(Elem x s) = multipopT k xs
    + phi (multipop k xs) - phi xs
-- Use definition of multipopT n (Elem _ s)
  === (1 + multipopT (k - 1) s)
    + phi (snd (multipop k xs)) - phi xs
-- Use definition of multipop n (Elem _ s)
  === (1 + multipopT (k - 1) s)
    + phi (snd (multipop (k - 1) s)) - phi xs
-- Use definition of phi (Elem _ s)
  === (1 + multipopT (k - 1) s)
    + phi (snd (multipop (k - 1) s))
    - (1 + phi s)
-- Remove parentheses
  === 1 + multipopT (k - 1) s
    + phi (snd (multipop (k - 1) s)) - 1 - phi s
-- 1s cancel out
  === multipopT (k - 1) s
    + phi (snd (multipop (k - 1) s)) - phi s
-- use induction hypothesis
  ? multipopP (k - 1) s
  *** QED
\end{lstlisting}

While this stack data structure might look contrived, it is basically half of Okasaki's queues~\cite{okasaki}. A queue is one front and one back list (or stack) where the back list is in reverse order. This allows us to remove elements from the front list and add elements to the back list in \O{1}. When the front list is empty, the back list is rotated and used as the new front list, an operation which is similar to \texttt{multipop} here.

\section{Case study: Binomial Heaps}\label{sec:binomal_heap}

\begin{figure}
\includegraphics[width=\columnwidth]{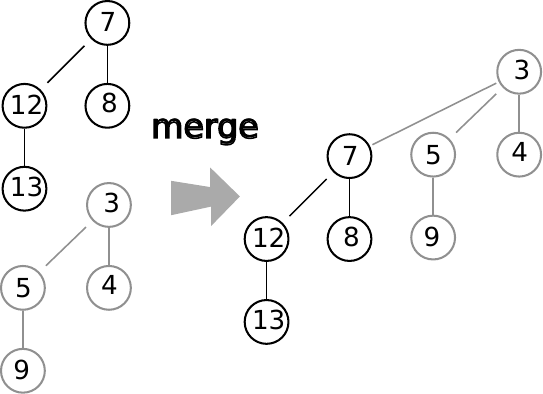}
\caption{Merge of two binomial trees of rank 2, adapted from Wikipedia~\cite{heap_merge}}\label{fig:heap_merge}
\end{figure}

In functional languages priority queues are usually implemented as some kind of (min) heap. A binomial heap is a forest of binomial trees where there exists at most one tree of every rank. A binomial tree of rank zero is just the root. A binomial tree of rank $k$ consists of its root and one binomial tree for every rank from $k - 1$ to zero as its children. Two binomial trees of rank $k$ can be merged by attaching the tree with the bigger root as first child of the other tree resulting in a binomial tree of rank $k + 1$ (see figure~\ref{fig:heap_merge}).

To insert a new element into a binomial heap, we insert a new tree of rank zero into the forest. If the forest already contains a tree of rank zero the \texttt{merge} operation is used to get a tree of rank one. If there is already a tree of rank one in the forest, they get merged again, and so on. Once a free spot is found, insertion is finished.

We can define the type of trees and of forest, as well as the operations like this in Haskell:

\noindent
\begin{minipage}{\linewidth}
\begin{lstlisting}
data Tree a = MkTree a [Tree a]

{-@ reflect mergeTree @-}
mergeTree :: Ord a =>
  Tree a -> Tree a -> Tree a
mergeTree l@(MkTree lr lc) r@(MkTree rr rc)
  | lr <= rr = MkTree lr (r : lc)
  | otherwise = MkTree rr (l : rc)

data Forest a = FEnd -- empty forest
  | F0 (Forest a) -- no tree at this position
  | F1 (Tree a) (Forest a)
type Heap a = Forest a

{-@ reflect insertTree @-}
insertTree :: Ord a =>
  Tree a -> Heap a -> Heap a
insertTree t FEnd = F1 t FEnd
insertTree t (F0 f) = F1 t f
insertTree t (F1 t' f) =
  F0 (insertTree (mergeTree t t') f)

\end{lstlisting}
\end{minipage}

Ideally we would use more advanced type level features of Haskell to ensure correctness of our implementation. For example we could ensure that the children of the root of a binomial tree have the correct ranks in descending order. However at the time of writing some language extensions necessary for this like \texttt{DataKinds} are not well supported by LiquidHaskell and cause problems in the proofs (see appendix~\ref{app:datakinds} for the definition of such a correct-by-construction binomial heap).

For the amortized analysis of the binomial heap, \texttt{merge} only does one comparison and one allocation no matter which rank the trees are of so it is in \O{1}. For the insertion we only work on the roots of which there are at most $log_2 \: n$ many.\footnote{A tree of rank $k$ consists of two trees of rank $k - 1$, so the number of elements in the tree always doubles} So the operation is in \O{log \: n}.

As our potential $\Phi$ we will use the number of trees in the forest. The intuition here is that the worst case happens when there is a tree at every position and we need to walk through the whole forest to find a free spot. This also fulfils the requirement that the empty data structure has a potential of zero. Again for clarity we will define the timing functions manually to show how they are derived from the original operations. As the merge operation is constant we include its cost directly in the insert operation for brevity. The proof is very similar to the \texttt{multipop} example from section~\ref{sec:stack}, so we will use \textit{ple} to automate most of it.

\begin{lstlisting}[escapechar=!]
{-@ reflect phi @-}
{-@ phi :: Heap a -> Nat @-}
phi :: Heap a -> Int
phi FEnd = 0
phi (F0 rest) = phi rest
phi (F1 rest) = 1 + phi rest

{-@ reflect insertT @-}
{-@ insertT :: Ord a !\color{codegreen}{=>}!
  Tree a -> Heap a -> { x:Int | x >= 1 } @-}
insertT :: Ord a =>
  Tree a -> Heap a -> Int
insertT _ FEnd = 1
insertT _ (F0 _) = 1
insertT t (F1 t' f) = 1 +
  insertT (mergeTree t t') f

{-@ automatic-instances insertTreeP @-}
{-@ insertTreeP :: t:Tree a -> f:Heap a ->
  { insertT t f + phi (insertTree t f)
    - phi f <= 2 }
@-}
insertTreeP :: Ord a => Tree a -> Heap a -> Proof
insertTreeP t (F1 t' f') = trivial
  ? insertTreeP (mergeTree t t') f'
insertTreeP _ _ = trivial
\end{lstlisting}

\section{Case study: Finger Trees}\label{sec:fingertree}

To show that our approach scales to more complicated data structures that are widely used in industry, we will prove the complexity of finger trees, the data structure behind \texttt{Data.Seq} from \textit{containers}~\cite{containers}. It was originally described by Hinze and Paterson~\cite{fingertrees}, but we will follow the simplified version of Claessen~\cite{fingertrees_new} here. The simplified version does not need nested pattern matches to implement concatenation of finger trees, which would otherwise lead to a combinatorial explosion in LiquidHaskell constraints.

\begin{lstlisting}
data Seq a = Nil
  | Unit a
  | More (Digit a) (Seq (Tuple a)) (Digit a)

data Digit a = One a | Two a a | Three a a a

data Tuple a = Pair a a | Triple a a a
\end{lstlisting}

The finger tree has between one and three elements on either side to enable cheap \texttt{cons} and \texttt{snoc}. The most interesting property of the data structure is the polymorphic recursion. It ensures an equal depth of nesting of \texttt{Tuple} types on both sides of the spine. It is also what might make it difficult to verify in other languages, as for example Isabelle/HOL~\cite{isabelle} does not directly support polymorphic recursion. In this case, it would be necessary to define the type such that it allows uneven nesting of \texttt{Tuple}s and then "carve out" the subset of valid finger trees. Other provers like Coq~\cite{coq} support polymorphic recursion, allowing a more direct one to one translation.

\subsection{Cons and Snoc}

Because the finger tree is a symmetric tree, \texttt{cons} and \texttt{snoc} are very similar. For this reason we show an explicit definition and proof only for \texttt{cons} and provide the definition and proof for \texttt{snoc} in the appendix. The base cases are straightforward and just use the flexibility of the digits to add the element to the front. The interesting case is when the front is already full and we need to recurse.

\begin{lstlisting}
{-@ reflect cons @-}
cons :: a -> Seq a -> Seq a
cons x Nil = Unit x
cons x (Unit y) = More (One x) Nil (One y)
cons x (More (One y) q u) = More (Two x y) q u
cons x (More (Two y z) q u) =
  More (Three x y z) q u
cons x (More (Three y z w) q u) =
  More (Two x y) (cons (Pair z w) q) u
\end{lstlisting}

The next step is to find a valid potential function $\Phi$. For this, we check which states of the data structure make \texttt{cons} expensive. As evident by the definition above, this is the case when the digit in the front is already full. So we can define the \textit{danger} of a digit as every state that will make the operation expensive. To also support \texttt{uncons} -- the inverse operation -- we not only mark a full digit as dangerous, but also those that have only one element inside (as this would make \texttt{uncons} go into the recursion). We then define the potential to be the sum of the danger in the tree. The timing functions are again directly derived from the \texttt{cons} operation itself. The proof itself can be mostly automated with \textit{ple}.

\begin{lstlisting}
{-@ reflect danger @-}
{-@ danger :: Digit a -> Nat @-}
danger :: Digit a -> Int
danger One{} = 1
danger Two{} = 0
danger Three{} = 1

{-@ reflect phi @-}
{-@ phi :: Seq a -> Nat @-}
phi :: Seq a -> Int
phi Nil = 0
phi Unit{} = 0
phi (More pr m sf) =
  danger pr + phi m + danger sf

{-@ reflect consT @-}
{-@ consT :: a -> Seq a -> { x:Int | x >= 1 } @-}
consT :: a -> Seq a -> Int
consT _ Nil = 1
consT _ (Unit _) = 1
consT _ (More (One _) _ _) = 1
consT _ (More (Two _ _) _ _) = 1
consT _ (More (Three _ z w) q _) =
  1 + consT (Pair z w) q

{-@ automatic-instances consP @-}
{-@ consP :: x:a -> t:Seq a ->
    { consT x t + phi (cons x t) - phi t <= 3 }
@-}
consP :: a -> Seq a -> Proof
consP _ (More (Three _ z w) q _) = trivial
  ? consP (Pair z w) q
consP _ _ = trivial
\end{lstlisting}

\subsection{Append}\label{sec:append}

Even though \texttt{append} has the same amortized complexity as its normal complexity, we need to show that it never increases the potential more than a logarithmic amount. Claessen~\cite{fingertrees_new} uses a helper function \texttt{glue} that concatenates two finger trees with at most three extra elements in the middle. Several other functions also require very specific lower and upper bounds on the length of their arguments. This can be directly expressed using LiquidHaskell.

\begin{lstlisting}
{-@ reflect glue @-}
{-@ glue :: Seq a -> { xs:[_] | len xs <= 3 }
    -> Seq a -> Seq a @-}
glue :: Seq a -> [a] -> Seq a -> Seq a
glue Nil as q2 = foldr cons q2 as
glue q1 as Nil = foldl snoc q1 as
glue (Unit x) as q2 = foldr cons q2 (x : as)
glue q1 as (Unit x) = snoc (foldl snoc q1 as) x
glue (More u1 q1 v1) as (More u2 q2 v2) =
  More u1 (glue q1 (
    toTuples (toList v1 ++ as ++ toList u2)
  ) q2) v2

{-@ reflect append @-}
append :: Seq a -> Seq a -> Seq a
append q1 q2 = glue q1 [] q2

{-@ reflect toList @-}
{-@ toList :: Digit a ->
  { xs:[a] | len xs >= 1 && len xs <= 3 } @-}
toList :: Digit a -> [a]
toList (One x) = [x]
toList (Two x y) = [x, y]
toList (Three x y z) = [x, y, z]

{-@ reflect toTuples @-}
{-@ toTuples ::
  { xs:[_] | len xs >= 2 && len xs <= 9 }
  -> { ys:[_] | len ys >= 1 && len ys <= 3 } @-}
toTuples :: [a] -> [Tuple a]
toTuples = toTuples'

{-@ reflect toTuples' @-}
{-@ toTuples' ::
  { xs:[_] | len xs != 1 && len xs <= 9 }
  -> { ys:[_] | if len xs == 0 then
        len ys == 0
      else if len xs <= 3 then
        len ys == 1
      else if len xs <= 6 then
        len ys == 2
      else len ys == 3 } @-}
toTuples' :: [a] -> [Tuple a]
toTuples' [] = []
toTuples' [x, y] = [Pair x y]
toTuples' [x, y, z, w] = [Pair x y, Pair z w]
toTuples' (x:y:z:xs) =
  Triple x y z : toTuples' xs
\end{lstlisting}

We also need to reimplement \texttt{foldl}, \texttt{foldr} and list concatenation because their definitions in the standard library do not come with \texttt{reflect} annotations so they cannot be used in a LiquidHaskell proof.

\begin{lstlisting}
{-@ reflect foldl @-}
foldl :: (b -> a -> b) -> b -> [a] -> b
foldl _ x [] = x
foldl f a (x:xs) = foldl f (f a x) xs

{-@ reflect foldr @-}
foldr :: (a -> b -> b) -> b -> [a] -> b
foldr _ x [] = x
foldr f a (x:xs) = f x (foldr f a xs)

{-@ reflect ++ @-}
{-@ (++) :: x:[_] -> y:[_] ->
  { z:[_] | len z == len x + len y } @-}
(++) :: [a] -> [a] -> [a]
[] ++ ys = ys
(x:xs) ++ ys = x : (xs ++ ys)
\end{lstlisting}

For the amortization proof we of course have to use the same potential as for \texttt{cons}/\texttt{snoc}, and the timing functions are directly derived from the \texttt{glue} operation itself and all other functions called by it. Note that the timing functions of the higher-order \texttt{foldl} and \texttt{foldr} functions take a timing function as input.

\begin{lstlisting}
{-@ reflect foldrT @-}
{-@ foldrT :: (a -> b -> b) ->
  (a -> b -> { x:Int | x >= 1 }) -> b ->
  [a] -> { x:Int | x >= 1 } @-}
foldrT :: (a -> b -> b) -> (a -> b -> Int)
  -> b -> [a] -> Int
foldrT _ _ _ [] = 1
foldrT f fT b (x:xs) = fT x (foldr f b xs)
  + foldrT f fT b xs

{-@ reflect foldlT @-}
{-@ foldlT :: (b -> a -> b) ->
  (b -> a -> { x:Int | x >= 1 }) -> b ->
  [a] -> { x:Int | x >= 1 } @-}
foldlT :: (b -> a -> b) -> (b -> a -> Int)
  -> b -> [a] -> Int
foldlT _ _ _ [] = 1
foldlT f fT a (x:xs) = fT a x
  + foldlT f fT (f a x) xs

{-@ reflect glueT @-}
{-@ glueT :: Seq a -> { as:[a] | len as <= 3 }
  -> Seq a -> { x:Int | x >= 1 } @-}
glueT :: Seq a -> [a] -> Seq a -> Int
glueT Nil as q2 = 1 + foldrT cons consT q2 as
glueT q1 as Nil = 1 + foldlT snoc snocT q1 as
glueT (Unit x) as q2 =
  1 + foldrT cons consT q2 (x : as)
glueT q1 as (Unit x) =
  1 + snocT (foldl snoc q1 as) x
    + foldlT snoc snocT q1 as
glueT (More _ q1 v1) as (More u2 q2 _) =
  1 + glueT q1 (
      toTuples (toList v1 ++ as ++ toList u2)
    ) q2
\end{lstlisting}

We also need a way to specify the logarithmic complexity, so we will define a logarithm function as well as a way to calculate the number of elements in a finger tree.

\begin{lstlisting}
{-@ reflect log2 @-}
{-@ log2 :: { x:Int | x >= 1 } -> Nat @-}
log2 :: Int -> Int
log2 1 = 0
log2 n = 1 + log2 (n `div` 2)

{-@ reflect tuplesToList @-}
{-@ tuplesToList :: x:[_] -> { y:[_] |
    len y <= 3 * len x && len y >= 2 * len x
  } @-}
tuplesToList :: [Tuple a] -> [a]
tuplesToList [] = []
tuplesToList (Pair a b:xs) = a:b:tuplesToList xs
tuplesToList (Triple a b c:xs) =
  a:b:c:tuplesToList xs

{-@ reflect seqToList @-}
seqToList :: Seq a -> [a]
seqToList Nil = []
seqToList (Unit x) = [x]
seqToList (More u q v) = toList u ++
  tuplesToList (seqToList q) ++ toList v
\end{lstlisting}

For the base cases of \texttt{glue} it is helpful to prove that folding \texttt{cons}/\texttt{snoc} over a list takes as many time steps as the amortized complexity of \texttt{cons}/\texttt{snoc} times the number of elements in the list.

\begin{lstlisting}
{-@ automatic-instances foldrTCons @-}
{-@ foldrTCons :: q:Seq a -> as:[a] ->
  { foldrT cons consT q as + pot (foldr cons q as)
    - pot q <= 3 * len as + 1 } @-}
foldrTCons :: Seq a -> [a] -> Proof
foldrTCons _ [] = trivial
foldrTCons q as@(x:xs) =
  foldrT cons consT q as + pot (foldr cons q as)
    - pot q
  === consT x (foldr cons q xs)
      + foldrT cons consT q xs
      + pot (cons x (foldr cons q xs)) - pot q
      + pot (foldr cons q xs)
      - pot (foldr cons q xs)
  ? consAmortized x (foldr cons q xs)
  =<= 3 + foldrT cons consT q xs
    + pot (foldr cons q xs) - pot q
  ? foldrTCons q xs
  =<= 3 + 3 * length xs + 1
  *** QED

{-@ automatic-instances foldlTSnoc @-}
{-@ foldlTSnoc :: q:Seq a -> as:[a] ->
  { foldlT snoc snocT q as
    + pot (foldl snoc q as)
    - pot q <= 3 * len as + 1 } @-}
foldlTSnoc :: Seq a -> [a] -> Proof
foldlTSnoc _ [] = trivial
foldlTSnoc q as@(x:xs) =
  foldlT snoc snocT q as + pot (foldl snoc q as)
    - pot q
  === snocT q x
      + foldlT snoc snocT (snoc q x) xs
      + pot (foldl snoc (snoc q x) xs) - pot q
      + pot (snoc q x) - pot (snoc q x)
  ? snocAmortized q x
  =<= 3 + foldlT snoc snocT (snoc q x) xs
    + pot (foldl snoc (snoc q x) xs)
    - pot (snoc q x)
  ? foldlTSnoc (snoc q x) xs
  *** QED
\end{lstlisting}

Now for the actual proof we will only show one of the base cases as the others follow the same schema. All the base cases have a constant upper bound on their complexity. For the recursive case we also need some small helper facts about the logarithm such as monotonicity. We also use a \texttt{where} binding to factor out common code and make the proof more readable.

\begin{lstlisting}
{-@ log2Mono :: { x:Int | 1 <= x } ->
  { y:Int | x <= y } -> { log2 x <= log2 y } @-}
log2Mono :: Int -> Int -> Proof
log2Mono 1 m = log2 1 <= log2 m *** QED
log2Mono n m = log2 n <= log2 m
  === 1 + log2 (n `div` 2)
    <= 1 + log2 (m `div` 2)
  === log2 (n `div` 2) <= log2 (m `div` 2)
  ? log2Mono (n `div` 2) (m `div` 2)
  *** QED

{-@ divCancel :: x:Int ->
  { div (2 * x) 2 == x } @-}
divCancel :: Int -> Proof
divCancel _ = trivial

{-@ glueAmortized :: q1:Seq a
  -> { as:[a] | len as <= 3 }
  -> q2:Seq a -> {
      glueT q1 as q2 + pot (glue q1 as q2)
      - pot q1 - pot q2
    <= log2 (max (len (seqToList q1)
      + len (seqToList q2)) 2) + 14
  } @-}
glueAmortized :: Seq a -> [a] -> Seq a -> Proof
glueAmortized q1@Nil as q2 =
  glueT q1 as q2 + pot (glue q1 as q2)
    - pot q1 - pot q2
  === 1 + foldrT cons consT q2 as
    + pot (foldr cons q2 as) - pot q1 - pot q2
  ? foldrTCons q2 as
  =<= 1 + 3 * length as + 1 - pot q1
  =<= log2 (max (length (seqToList q1)
    + length (seqToList q2)) 2) + 14
  *** QED
glueAmortized qq1@(More u1 q1 v1) as
  qq2@(More u2 q2 v2) =
  glueT qq1 as qq2 + pot (glue qq1 as qq2)
  - pot qq1 - pot qq2
  === 1 + glueT q1 (toTuples
        (toList v1 ++ as ++ toList u2)) q2
      + pot (More u1 (glue q1 (toTuples
          (toList v1 ++ as ++ toList u2)) q2) v2)
      - danger u1 - pot q1 - danger v1 - danger u2
      - pot q2 - danger v2
  === 1 + glueT q1 (toTuples
          (toList v1 ++ as ++ toList u2)) q2
      + pot (glue q1 (toTuples
          (toList v1 ++ as ++ toList u2)) q2)
      - pot q1 - danger v1 - danger u2 - pot q2
-- apply induction
  ? glueAmortized q1
    (toTuples (toList v1 ++ as ++ toList u2)) q2
  =<= log2 (max n 2) + 15
-- move (+1) into the logarithm
  ? divCancel (max n 2)
  === 1 + log2 (2 * max n 2 `div` 2) + 14
  === log2 (2 * max n 2) + 14

-- simplify using monotonicity of log2
  ? log2Mono (2 * max n 2) (4 + 2 * n1 + 2 * n2)
  =<= log2 (4 + 2 * n1 + 2 * n2) + 14
  ? log2Mono (4 + 2 * n1 + 2 * n2) (4 + m1 + m2)
  =<= log2 (4 + m1 + m2) + 14
  ? log2Mono (4 + m1 + m2)
    (max (length (toList u1
      ++ tuplesToList (seqToList q1) ++ toList v1
    ) + length (toList u2
      ++ tuplesToList (seqToList q2) ++ toList v2
    )) 2)
  =<= log2 (max (length (toList u1
      ++ tuplesToList (seqToList q1) ++ toList v1
    ) + length (toList u2
      ++ tuplesToList (seqToList q2) ++ toList v2
    )) 2) + 14
  === log2 (max (length (seqToList qq1)
    + length (seqToList qq2)) 2) + 14
  *** QED
  where
    n1 = length (seqToList q1)
    n2 = length (seqToList q2)
    n = n1 + n2
    m1 = length (tuplesToList (seqToList q1))
    m2 = length (tuplesToList (seqToList q2))
\end{lstlisting}

\subsection{Comparison with Claessen's formalization}

For their formalization, Claessen used HipSpec~\cite{hipspec}, a tool to translate Haskell to first order logic. However we were not able to directly compare these proofs to ours as the tool has not seen any updates in the last 5 years and the last supported GHC version is 7.8. LiquidHaskell on the other hand has seen some industry adoption and has kept up to date with new GHC versions (version 9.8 at the time of writing).

Aside from these technical limitations, they only formalized the comparatively easy \texttt{cons} and \texttt{snoc} complexities and leave out the much more interesting logarithmic complexity of append.

\section{Conclusion}\label{sec:conclusion}

In this paper we successfully applied LiquidHaskell to prove the amortized complexity of increasingly complex data structures. Despite some limitations of LiquidHaskell, like incomplete support for \texttt{DataKinds}, proving amortized complexity directly in Haskell is quite feasible. However some more complex proofs might require a lot of code for rather simple proof steps (compare the use of monotonicity in the previous section). In such cases, the extra automation built into modern theorem provers would certainly improve the user experience.

On the other hand, LiquidHaskell's explicit equational reasoning is easier to follow for a new learner than opaque proof scripts written in a tactic-based theorem prover like Coq or Lean\footnote{Lean does support equational reasoning via the \texttt{calc} environment, however most proofs are done using tactic scripts}. While Isabelle/HOL also features equational reasoning with Isar~\cite{isar}, it does not support polymorphic recursion which makes the definition of our third case study a lot harder.

For future work, it would be interesting to investigate how complex an adaptation of our technique for reasoning about lazy semantics would be. This would allow to expand the complexity results to a shared setting where reuse of intermediate results is allowed. It might be possible to adapt \citet{forking_paths}'s or \citet{lazyfunction}'s works on complexity for lazy programs, but LiquidHaskell's lack of quantifiers will make this challenging.


\bibliographystyle{ACM-Reference-Format}
\bibliography{references}

\appendix

\section{Correct-by-construction binomial heap}\label{app:datakinds}

\begin{lstlisting}
import Data.Kind (Type)

data Nat = Zero | Succ Nat

type Tree :: Nat -> Type -> Type
data Tree k a where
  MkTree :: a -> DecrList Tree k a -> Tree k a

type DecrList :: (Nat -> Type -> Type)
  -> Nat -> Type -> Type
data DecrList t k a where
  DNil :: DecrList t Zero a
  DCons :: t k a -> DecrList t k a
    -> DecrList t (Succ k) a

mergeTree :: Ord a => Tree k a -> Tree k a
  -> Tree (Succ k) a
mergeTree l@(MkTree lr lc) r@(MkTree rr rc)
  | lr <= rr = MkTree lr (DCons r lc)
  | otherwise = MkTree rr (DCons l rc)

data Binary
  = B0 Binary
  | B1 Binary
  | BEnd

type BInc :: Binary -> Binary
type family BInc b where
  BInc BEnd = B1 BEnd
  BInc (B0 rest) = B1 rest
  BInc (B1 rest) = B0 (BInc rest)

type Forest :: Nat -> Binary -> Type -> Type
data Forest k b a where
  FEnd :: Forest k BEnd a
  F0 :: Forest (Succ k) b a -> Forest k (B0 b) a
  F1 :: Tree k a -> Forest (Succ k) b a
    -> Forest k (B1 b) a

type Heap :: Binary -> Type -> Type
newtype Heap b a = MkHeap (Forest Zero b a)

insert :: Ord a => Tree k a -> Forest k b a
  -> Forest k (BInc b) a
insert t FEnd = F1 t FEnd
insert t (F0 f) = F1 t f
insert t (F1 t' f) =
  F0 (insert (mergeTree t t') f)
\end{lstlisting}

\balance

\section{Definition and proof for the snoc operation on a finger tree}

\begin{lstlisting}
{-@ reflect snoc @-}
snoc :: Seq a -> a -> Seq a
snoc Nil x = Unit x
snoc (Unit x) y = More (One x) Nil (One y)
snoc (More u q (One x)) y = More u q (Two x y)
snoc (More u q (Two x y)) z =
  More u q (Three x y z)
snoc (More u q (Three x y z)) w =
  More u (snoc q (Pair x y)) (Two z w)

{-@ reflect snocT @-}
{-@ snocT :: Seq a -> a -> { x:Int | x >= 1 } @-}
snocT :: Seq a -> a -> Int
snocT Nil _ = 1
snocT (Unit _) _ = 1
snocT (More _ _ (One _)) _ = 1
snocT (More _ _ (Two _ _)) _ = 1
snocT (More _ q (Three x y _)) _ =
  1 + snocT q (Pair x y)


{-@ automatic-instances snocAmortized @-}
{-@ snocAmortized :: q:Seq a -> x:a ->
  { snocT q x + pot (snoc q x) - pot q <= 3 } @-}
snocAmortized :: Seq a -> a -> Proof
snocAmortized (More _ q (Three x y _)) _ =
  trivial ? snocAmortized q (Pair x y)
snocAmortized _ _ = trivial

\end{lstlisting}

\end{document}